\documentclass[aps,prl,twocolumn,superscriptaddress,showpacs]{revtex4}
 
\usepackage{graphicx}

\begin{document}
 
\preprint{XXXX-XX}

\title{Longitudinal-Transverse Separations of 
Structure Functions at Low $Q^{2}$ for Hydrogen and Deuterium}
 
\author{V.~Tvaskis}
\affiliation{Vrije Universiteit, 1081 HV Amsterdam, The Netherlands}
\affiliation{National Instituut voor Kernfysica en Hoge-Energiefysica (NIKHEF), 
1009 DB Amsterdam, The Netherlands}
\author{M.~E.~Christy}
\affiliation{Hampton University, Hampton, Virginia 23668}

\author{J.~Arrington}
\affiliation{Argonne National Laboratory, Argonne, Illinois 60439}

\author{R.~Asaturyan}
\affiliation{Yerevan Physics Institute, 375036, Yerevan, Armenia}

\author{O.~K.~Baker}
\affiliation{Hampton University, Hampton, Virginia 23668}

\author{H.~P.~Blok}
\affiliation{Vrije Universiteit, 1081 HV Amsterdam, The Netherlands}
\affiliation{National Instituut voor Kernfysica en Hoge-Energiefysica (NIKHEF),
1009 DB Amsterdam, The Netherlands}

\author{P.~Bosted}
\affiliation{Thomas Jefferson National Accelerator Facility, Newport News, Virginia 23606}

\author{M.~Boswell}
\affiliation{Randolph-Macon Woman's College, Lynchburg, Virginia 24503}

\author{A.~Bruell}
\affiliation{Massachusetts Institute of Technology, Cambridge, Massachusetts 02139}

\author{A.~Cochran}
\affiliation{Hampton University, Hampton, Virginia 23668}

\author{L.~Cole}
\affiliation{Hampton University, Hampton, Virginia 23668}

\author{J.~Crowder}
\affiliation{Juniata College, Huntingdon, Pennsylvania 16652}

\author{J.~Dunne}
\affiliation{Mississippi State University, Mississippi State, Mississippi 39762}

\author{R.~Ent}
\affiliation{Thomas Jefferson National Accelerator Facility, Newport News, Virginia 23606}

\author{H.~C.~Fenker}
\affiliation{Thomas Jefferson National Accelerator Facility, Newport News, Virginia 23606}

\author{B.~W.~Filippone}
\affiliation{California Institute of Technology, Pasadena, California 91125}

\author{K.~Garrow}
\affiliation{Thomas Jefferson National Accelerator Facility, Newport News, Virginia 23606}

\author{A.~Gasparian}
\affiliation{Hampton University, Hampton, Virginia 23668}

\author{J.~Gomez}
\affiliation{Thomas Jefferson National Accelerator Facility, Newport News, Virginia 23606}

\author{H.E.~Jackson}
\affiliation{Argonne National Laboratory, Argonne, Illinois 60439}

\author{C.~E.~Keppel}
\affiliation{Hampton University, Hampton, Virginia 23668}
\affiliation{Thomas Jefferson National Accelerator Facility, Newport News, Virginia 23606}

\author{E.~Kinney}
\affiliation{University of Colorado, Boulder, Colorado 80309}

\author{Y.~Liang}
\affiliation{Hampton University, Hampton, Virginia 23668}
\affiliation{American University, Washington, D.C. 20016}

\author{W.~Lorenzon}
\affiliation{University of Michigan, Ann Arbor, Michigan 48109}

\author{A.~Lung}
\affiliation{Thomas Jefferson National Accelerator Facility, Newport News, Virginia 23606}

\author{D.~J.~Mack}
\affiliation{Thomas Jefferson National Accelerator Facility, Newport News, Virginia 23606}

\author{J.~W.~Martin}
\affiliation{Massachusetts Institute of Technology, Cambridge, Massachusetts 02139}

\author{K.~McIlhany}
\affiliation{Massachusetts Institute of Technology, Cambridge, Massachusetts 02139}

\author{D.~Meekins}
\affiliation{Thomas Jefferson National Accelerator Facility, Newport News, Virginia 23606}

\author{R.~G.~Milner}
\affiliation{Massachusetts Institute of Technology, Cambridge, Massachusetts 02139}

\author{J.~H.~Mitchell}
\affiliation{Thomas Jefferson National Accelerator Facility, Newport News, Virginia 23606}

\author{H.~Mkrtchyan}
\affiliation{Yerevan Physics Institute, 375036, Yerevan, Armenia}

\author{B.~Moreland}
\affiliation{Thomas Jefferson National Accelerator Facility, Newport News, Virginia 23606}

\author{V.~Nazaryan}
\affiliation{Hampton University, Hampton, Virginia 23668}

\author{I.~Niculescu}
\affiliation{The George Washington University, Washington, D.C. 20052}

\author{A.~Opper}
\affiliation{Ohio University, Athens, Ohio 45071}

\author{R.~B.~Piercey}
\affiliation{Mississippi State University, Mississippi State, Mississippi 39762}

\author{D.H.~Potterveld}
\affiliation{Argonne National Laboratory, Argonne, Illinois 60439}

\author{B.~Rose}
\affiliation{Thomas Jefferson National Accelerator Facility, Newport News, Virginia 23606}

\author{Y.~Sato}
\affiliation{Hampton University, Hampton, Virginia 23668}

\author{W.~Seo}
\affiliation{Kyungpook National University, Taegu 702-701, Korea}

\author{G.~Smith}
\affiliation{Thomas Jefferson National Accelerator Facility, Newport News, Virginia 23606}

\author{K.~Spurlock}
\affiliation{Mississippi State University, Mississippi State, Mississippi 39762}

\author{G.~van der Steenhoven}
\affiliation{National Instituut voor Kernfysica en Hoge-Energiefysica (NIKHEF), 
1009 DB Amsterdam, The Netherlands}

\author{S.~Stepanyan}
\affiliation{Yerevan Physics Institute, 375036, Yerevan, Armenia}

\author{V.~Tadevosian}
\affiliation{Yerevan Physics Institute, 375036, Yerevan, Armenia}

\author{A.~Uzzle}
\affiliation{Hampton University, Hampton, Virginia 23668}

\author{W.~F.~Vulcan}
\affiliation{Thomas Jefferson National Accelerator Facility, Newport News, Virginia 23606}

\author{S.~A.~Wood}
\affiliation{Thomas Jefferson National Accelerator Facility, Newport News, Virginia 23606}

\author{B.~Zihlmann}
\affiliation{Vrije Universiteit, 1081 HV Amsterdam, The Netherlands}

\author{V.~Ziskin}
\affiliation{Massachusetts Institute of Technology, Cambridge, Massachusetts 02139}

\date{\today}

\begin{abstract}

We report on a study of the longitudinal
to transverse cross section ratio, $R=\sigma_L/\sigma_T$, 
at low values of $x$ and $Q^{2}$,
as determined from inclusive inelastic electron-hydrogen and electron-deuterium 
scattering data from Jefferson Lab Hall C spanning the four-momentum transfer 
range 0.06 $ < Q^{2} < 2.8$ GeV$^{2}$. 
Even at the lowest values of $Q^{2}$, $R$ remains nearly constant
and does not disappear with decreasing $Q^{2}$, as expected.
We find a nearly identical behaviour for hydrogen and deuterium. 

\end{abstract}


\maketitle

Since the early experiments at the Stanford Linear Accelerator Center (SLAC),
which discovered the substructure of the nucleon and led to the development of the
quark parton model \cite{parton},
deep inelastic scattering (DIS) has been a powerful tool
in the investigation of the partonic substructure of the nucleon.
After decades of experiments with electron and muon 
beams, the nucleon structure function $F_{2}(x, Q^{2})$ is known with high
precision over many orders of magnitude in $x$ and $Q^{2}$ \cite{pdg}.
Here, $Q^{2}$ is the negative square of the four-momentum transfer
of the exchanged virtual photon in the scattering process.
The Bjorken scaling variable $x=Q^2/2M\nu$, with $M$ the nucleon mass and $\nu$ the energy
transfer, can be interpreted as 
the fraction of the nucleon momentum carried by the struck parton.

In the region of large $Q^{2}$ and $\nu$, the results of
DIS measurements are typically interpreted in terms of 
partons (quarks and gluons).
In this case, the theoretical framework is provided by perturbative
Quantum Chromo Dynamics (pQCD), which includes logarithmic scaling violations.
This description starts to fail when non-perturbative effects such as
higher-twist interactions between the struck quark in the scattering process 
and other quarks or gluons in the nucleon
become important. 
The sensitivity for higher twist effects increases 
with decreasing $Q^{2}$, since they are proportional to powers of $1/Q^2$.
 
There is great interest in the behaviour of the nucleon structure functions
in the low $Q^{2}$ region where the transition 
from perturbative to non-perturbative QCD takes place. However,
little is known about this behaviour, since at large invariant mass $W$ of the hadronic system there are few data points 
in this region,
except for the (transverse) cross section $\sigma_{T}$ at exactly $Q^{2}$ = 0, which is 
accessible through real photon absorption experiments.
The more plentiful data at low $W$ are typically interpreted in terms of nucleon resonance excitations.

The differential cross section for inclusive electron scattering, after dividing by
the virtual photon flux factor ($\Gamma$), can be written as
\begin{equation}
\label{sigma}
{1 \over \Gamma }{ d^{2}\sigma \over d\Omega dE' } = \sigma_{T} + \varepsilon \sigma_{L} \label{eq:sigma} \phantom{l}, 
\end{equation}
where $\varepsilon$ is the virtual photon polarization and 
$\sigma_{L}$ ($\sigma_{T}$) is the longitudinal (transverse) virtual-photon
absorption cross section, which depends on $x$ and $Q^2$. 
Current conservation determines the behaviour of the structure functions 
for $Q^{2} \to$ 0, leading to
\begin{equation}
\label{Rdef}
R(x,Q^2) \equiv {\sigma_L\over\sigma_T} = { F_L\over{2xF_1} } = {\mathcal O}(Q^{2}),
\end{equation}
where $F_L$ and $F_1$ are the longitudinal and transverse nucleon structure 
functions.
The value of $Q^{2}$ at which this behavior
becomes manifest is however neither predictable nor yet observed.
 
While there is a wealth of data for $F_2=(2xF_1+F_{L})/(1+ Q^{2} / \nu^{2})$, 
relatively few data exist for $F_L$, or equivalently $R$.
Data on $R(x, Q^{2})$ on hydrogen and deuterium are available in a limited $x$
and $Q^{2}$ ($Q^{2}$ $>$ 1 GeV$^2$) range, with a typical uncertainty of
0.1-0.2 \cite{r1990,rbcdms}, comparable to the size of $R$ itself. 
For scattering from point-like spin-1/2 particles, $R$ should 
vanish at large values of $Q^{2}$ 
because of helicity conservation. At low values of $Q^{2}$, however,
$R$ is not small, and typical values are about 0.3.

Precision data on $R$ are necessary for several reasons.
Most importantly, determinations of the structure function $F_2$ 
from cross section measurements, and the parton distributions derived therefrom,
need numerical values for $R$. If the former are not based on longitudinally 
and transversely  separated measurements,
the uncertainties in $F_{2}$ introduced by assumptions for $R$
can be as large as 20\%.
Furthermore, especially in the low $Q^2$ region, data are needed to study 
higher twist effects and to search for the onset of the current
conservation behaviour of the structure functions at low $Q^2$ described above.

The determination of $R$ is typically accomplished via a Rosenbluth-type 
separation technique, which requires high precision measurements 
of the cross section at fixed values of $x$ and $Q^{2}$, but at different
values of $\varepsilon$. This technique requires the use of
at least two beam energies and correspondingly different scattered-electron angles, $\theta$.
Only in some experiments have such measurements 
actually been performed~\cite{r1990,rbcdms,dasu3,eric,yl1,remc,rnmc,tao}.

In the framework of perturbative QCD, there is no requirement that 
$R$ be the same for protons and neutrons. 
Previous results have shown $R^{D}$ - $R^{H}$, the difference in $R$ 
from hydrogen and deuterium targets, 
to be consistent with zero~\cite{r1990,rnmc,tao}.
However, these measurements were carried out mainly at higher $Q^2$,
where $R$ itself is small and any difference, therefore, difficult
to ascertain.
  
In this paper we present results from  a study of $R$ for
both hydrogen and deuterium  at low values of $Q^{2}$.
The experiment (E99-118) was carried out at the Thomas Jefferson National Accelerator Facility (Jlab).
Data were obtained at 0.007$ < x < $ 0.55 and
0.06 $ < Q^{2} < $ 2.8 GeV$^{2}$, by utilizing 2.301, 3.419 and 5.648 GeV
electron beams at a current of $I$ = 25 $\mu$A. The minimum scattered electron 
energy was $E' \approx 0.4$ GeV and the range of the 
invariant mass squared of the hadronic system $W^2$ was between 3.5 and 10 GeV$^2$. 
Electrons scattered from 4 cm long liquid hydrogen (H) and deuterium (D) targets
were detected in the High-Momentum  Spectrometer (HMS) in Hall C
at various angles between 10$^{\circ}$ and 60$^{\circ}$ . 
                                                                                   
The inclusive double differential cross section for each energy and 
angle bin within the spectrometer acceptance was determined from
\begin{equation}
\frac {d \sigma} {d \Omega dE^{'}} = 
\frac{Y_{corr}}{L \Delta \Omega \Delta E^{'}},
\end{equation}
\noindent where 
$\Delta \Omega$ ($\Delta E'$) is the bin width in solid angle 
(scattered energy), $L$ is 
the total integrated luminosity, and  $Y_{corr}$ is the measured
electron yield after correcting for detector 
inefficiencies, background events, and radiative effects. 

To account for backgrounds
from $\pi^0$ production and its subsequent decay into two photons followed 
by pair production of  
electron-positron pairs, positron data were also taken
by reversing the polarity of the spectrometer.
Other background contributions include electron scattering from the aluminum walls of the
cryogenic target cells and electroproduced
negatively charged pions. 
Events from the former were subtracted by performing substitute empty target
runs, while events from the latter were identified and
removed by use of both a gas Cherenkov counter
and an electromagnetic calorimeter. Additional details regarding the
analysis and the standard Hall
C apparatus employed in this experiment can be found in Ref. \cite{tvaskis}. 

Radiative effects including bremsstrahlung, vertex corrections and loop diagrams
are calculated using the approach by Bardin et al \cite{bardin}.
Additional radiative effects in the target and its exit windows were determined using 
the formalism of Mo and Tsai \cite{motsai}. 
The calculation of such effects includes the emission of one hard photon.
There is, however, the possibility that the electron could emit two hard photons. 
The calculation of this process is unfortunately not fully established and the 
corresponding effect was therefore treated in the present 
analysis as a separate uncertainty.

For each energy bin, a weighted average cross section 
over $\theta$ within the spectrometer acceptance was obtained
by using a model to correct 
for the angular variation of the cross section from the central angle value.  
To minimize the dependence on the model used to compute both this 
correction and
the radiative effects, 
an iterative procedure was employed. 

The total systematic uncertainty in the differential cross section was taken as the quadratic sum
of all the systematic uncertainties contributing to the cross section measurement.
In a Rosenbluth separation one needs to distinguish between uncertainties 
that are correlated between measurements at different $\varepsilon$,
such as uncertainties in target thickness and integrated charge,
and uncorrelated ones.
Not including the contributions from radiative corrections,
the uncorrelated systematic uncertainties 
on the cross section measurements in this experiment amounted to 0.9\%, 
while the total systematic uncertainty was 1.35\%.

The size, and consequently the uncertainty, of the radiative effects strongly depends 
on the kinematics and is largest at low values of $E'$ where the measured cross 
section is dominated by events from elastic or quasi-elastic scattering with the emission of 
one or more photons in the initial or final state.
The estimate of these uncertainties was determined by varying all relevant input cross sections 
within their uncertainties, and amounted to as much as 1.5\% 
for hydrogen and 8.5\% for deuterium in the most extreme cases 
considered. It rapidly decreased for higher values of $E'$, i.e. higher 
values of $x$ and $Q^2$.
The much larger uncertainty in the deuterium cross section is due to the 
contribution from quasielastic scattering which can only be 
modelled approximately. 

The extractions of purely longitudinal and transverse cross sections and  
structure functions were accomplished via 
the Rosenbluth technique, where 
the reduced cross section  
is fit linearly as a function of $\varepsilon$, as in Eq.(\ref{eq:sigma}).
The intercept of such a fit gives the 
transverse cross section
$\sigma_T$ (and therefore the structure function $F_1(x,Q^2)$), while the 
slope gives the longitudinal cross section $\sigma_L$, from which 
the structure functions $R(x,Q^2)$ or $F_L(x,Q^2)$ can be extracted.  

\begin{figure}
\includegraphics[width=3.4in,bb=0 0 550 517]{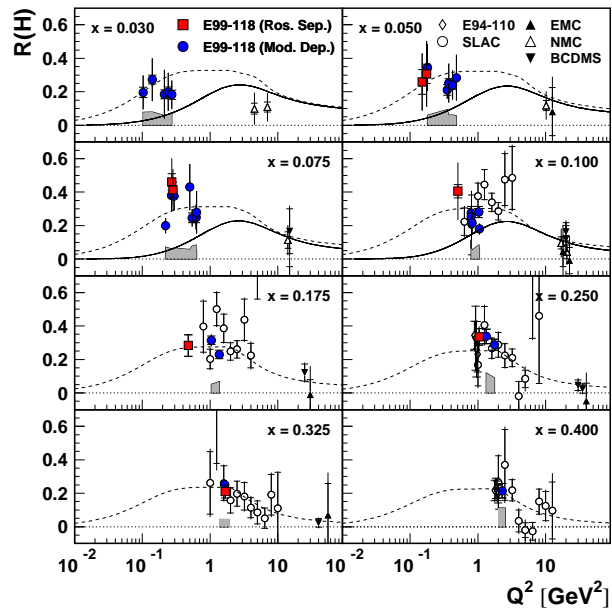}
\caption{\label{rvsw2}Comparison of the values of $R(x, Q^{2})$ for hydrogen from the 
present experiment (E99-118) to the results of other experiments. 
The dashed curve represents the parametrization $R^{H}_{e99118}(x, Q^{2})$ 
and the solid curve represents the model developed 
in \protect \cite{badelek} (see text for details.)}
\end{figure}

The results for $R(x,Q^2)$ in hydrogen
are shown in Fig. \ref{rvsw2} as a function of $Q^2$ 
for fixed values of $x$ (red squares). 
The inner error bars represent the combined statistical and uncorrelated systematic uncertainties.
The total error bars represent the statistical and systematic uncertainties added in quadrature. 
The data are compared to the results of previous measurements at Jlab (E94-110)~\cite{yl1}, SLAC~\cite{r1990}, 
and by the EMC~\cite{remc}, NMC~\cite{rnmc} and BCDMS~\cite{rbcdms} collaborations at CERN.

The structure function $F_{2}$ determined through the 
Rosenbluth separation technique was found to 
agree to better than 2\% with a Regge-motivated parameterization of all previously
available deep-inelastic scattering data ~\cite{f2all}, 
even at very low values of $Q^{2}$ \cite{tvaskis}.
This facilitated the utilization of an alternative approach
where $R$ was calculated 
using this parametrization (including a 2\% uncertainty) for the structure function $F_{2}(x, Q^{2})$ from

\begin{equation}
{ d^{2}\sigma \over d\Omega dE' } = 
\sigma_{Mott}{2MxF_{2} \over Q^{2}\varepsilon}\Bigg({1+\varepsilon R \over 1+R}\Bigg) \label{rmod} \phantom{l}.
\end{equation}

The results (shown by the blue circles in Fig. \ref{rvsw2}) cover a larger kinematic range 
than the results from the Rosenbluth separation, since the measurement of this
experiment is combined via the model with results of previous experiments at different energies. 
The inner error bars represent the statistical uncertainty and the total error bars
represent the quadratic sum of the statistical uncertainty and the uncertainty 
in the radiative corrections 
due to the possible emission of two hard 
photons. 
The shaded bands represent the uncertainties in the radiative effects due to
uncertainties in the input cross sections. 
Especially in the model dependent extraction the uncertainties are dominated 
by the unceratinties in the radiative corrections which are correlated 
between data points. 
The results from the second method agree very well with those obtained from the
Rosenbluth separation method.
Good agreement is also found with previous experiments 
in the regions of $x$ and $Q^2$ where the data overlap.
 
As mentioned above, at low values of $Q^{2}$, current conservation requires 
$R$ to be proportional to $Q^{2}$.
However, in the data from the present experiment this behaviour is not yet
observed, and $R$ remains nearly constant over the measured range in $Q^2$.
Thus, the transition to
$R \propto Q^2$ must occur below a $Q^2$ value of about 0.1 GeV$^{2}$ 
at low $x$, or below 1 GeV$^2$ for $x > 0.2$.
This result will have a direct impact on structure function extractions at low $Q^2$.

The dashed curve in Fig. \ref{rvsw2} represents a new 
parametrization of $R$ ($R^{H}_{e99118}(x, Q^{2})$ \cite{tvaskis}) 
based on all available data including those from this experiment. 
The functional form of the parametrization has been chosen 
to satisfy the 
condition that $R$ vanishes as $Q^{2}$ goes to zero. At $Q^{2}$ = 2 GeV$^2$, it 
was connected to a previously obtained parametrization from SLAC~\cite{r1990} 
that is based on measurements at higher values of  $Q^{2}$.
The solid curves in the upper four panels of Fig. \ref{rvsw2} show,
within its range of applicability ($x \leq 0.1$),
the model developed in \cite{badelek}, which is based on the photon-gluon 
fusion mechanism and extrapolated into the region of low $Q^2$.

Both the Rosenbluth separation and the model dependent extraction
of $R$ were also carried out for the deuterium data. While the precision
of the results from the Rosenbluth separation is comparable to that of 
the hydrogen data, the systematic uncertainty in the model
dependent extraction is much bigger for the deuterium data
due to a large uncertainty in the calculation of the quasi-elastic
radiation tail which is significant at low $x$ and $Q^{2}$.
\begin{figure}
\includegraphics[width=3.4in,bb=15 15 280 262]{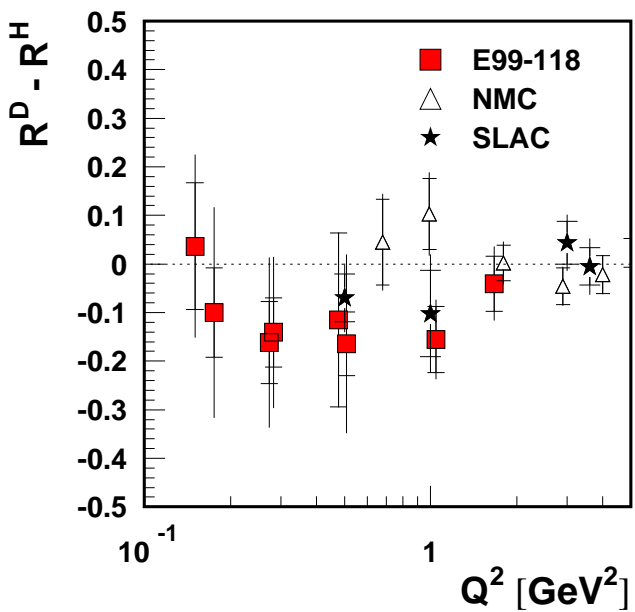}
\caption{\label{difdhr}The difference $R^D - R^H$ as a function of $Q^{2}$ from the present experiment 
calculated via Rosenbluth separation.
The data from previous experiments are also shown for $Q^2 < $ 5 GeV$^{2}$.}
\end{figure}  

Thus, the difference  $R^{D} - R^{H}$ was calculated using only the results from the
Rosenbluth separation method, compiled in Table~\ref{ros_sep}, and compared in Fig.~\ref{difdhr} 
to previous results from NMC~\cite{rnmc} and SLAC~\cite{tao}.
In this plot only the most recent and precise data from SLAC~\cite{tao} are shown, 
while additional SLAC results~\cite{r1990} are included in the statistical analysis 
of $R^{D} - R^{H}$, below.
Previously, the conclusion was drawn that there is no difference between $R^D$ and $R^H$.
However, most of the data from NMC are at rather high $Q^{2}$ values,
where $R$ itself is small, and the $R^{D} - R^{H}$ values extracted
from the SLAC measurements~\cite{r1990} were averaged over all $Q^{2}$, including
high $Q^{2}$ values, and were hence biased towards smaller differences.
Including our results, the data are still consistent with $R^D$ being identical to $R^H$. 
However, at values of $Q^2 < 1.5$ GeV$^2$  there is a hint both in the
present data and in the highest-precision data from SLAC~\cite{tao} that
$R^D$ is smaller than $R^H$. The global average (including all data) yields
$R^{D} - R^{H}$ = -0.054 $\pm$ 0.029. 

The results presented here are measurements of the 
longitudinal to transverse cross section ratio below $Q^{2}$ of about 
2 GeV$^2$
for hydrogen and deuterium targets. These data appear in a region where 
$R$ was expected to disappear as $Q^2$ gets very small. 
However, 
a nearly constant behaviour of $R^H$ and $R^D$ is observed down to
Q$^{2}$ of about 0.1 GeV$^{2}$ at low values of $x$. 
For $Q^2 <$ 1.5 GeV$^2$, the data 
hint at a small difference between $R^{D}$ and $R^{H}$.
    
\begin{acknowledgments}
This work is supported in
part by research grants from U.S. Department of Energy, the U.S. National Science
Foundation, and the Stichting voor Fundamenteel Oderzoek der Materie (FOM) of 
the Netherlands.
The Southeastern Universities Research Association operates the Thomas Jefferson National
Accelerator Facility under the U.S. Department of Energy contract 
DEAC05-84ER40150.
\end{acknowledgments}


\begin{table}
\caption{\label{ros_sep}The values of $R_H$ and $R^{D}$ - $R^{H}$ calculated via the Rosenbluth separation. 
Note that the systematic error in the difference accounts for the correlation between the uncertainties in the hydrogen and deuterium 
data. Complete data tables and the new parametrization of $R(x,Q^{2})$ may be requested via 
email (tvaskis@jlab.org, bruell@jlab.org).}
\begin{tabular}{c c c c c c c c}
$Q^{2}$ GeV$^2$  & $x$      & $R^{H}$ & Stat.  &  Syst.  & $R^{D}$ - $R^{H}$	& Stat. &  Syst.  \\
\hline
0.150       	 &  0.041   & 0.259   & 0.074  & 0.153   & 	  0.036 	& 0.131 & 0.136  \\
0.175       	 &  0.050   & 0.307   & 0.056  & 0.188   & 	 -0.100 	& 0.091 & 0.196  \\
0.273       	 &  0.077   & 0.460   & 0.049  & 0.132   & 	 -0.162 	& 0.084 & 0.153  \\
0.283       	 &  0.081   & 0.414   & 0.045  & 0.117   & 	 -0.141 	& 0.071 & 0.138  \\
0.476       	 &  0.156   & 0.283   & 0.063  & 0.025   & 	 -0.115 	& 0.179 & 0.021  \\
0.508       	 &  0.091   & 0.406   & 0.038  & 0.168   & 	 -0.164 	& 0.065 & 0.172  \\
1.045       	 &  0.200   & 0.335   & 0.048  & 0.041   & 	 -0.155 	& 0.068 & 0.046  \\
1.670       	 &  0.320   & 0.211   & 0.038  & 0.021   & 	 -0.040 	& 0.057 & 0.051  \\
\end{tabular}
\end{table}


\vspace*{-0.5cm}
\begin{thebibliography}{99}
\vspace*{-0.5cm}
\bibitem{parton} R. P. Feynman, {\it Photo-Hadron Interactions}, W. A. Benjamin Inc. Reading, MA (1976).
\bibitem{pdg} K. Hagiwara {\it et al.}, Phys. Rev. D {\bf 66}, 010001 (2002).
\bibitem{r1990} L. W. Whitlow {\it et al.}, Phys. Lett. {\bf B250}, 193 (1990). 
\bibitem{rbcdms} A. C. Benvenuti {\it et al.}, Phys. Lett. {\bf B223}, 485 (1989).
\bibitem{dasu3} S. Dasu {\it et al.}, Phys. Rev. Lett. {\bf 60}, 2591 (1988). 
\bibitem{eric} M. E. Christy {\it et al.}, Phys. Rev. C {\bf 70}, 015206 (2004).
\bibitem{yl1} Y. Liang {\it et al.}, nucl-ex/0410027 (2004).
\bibitem{remc} J. J. Aubert {\it et al.}, Nucl. Phys. {\bf B259}, 189 (1985).
\bibitem{rnmc} M. Arneodo {\it et al.}, Nucl. Phys. {\bf B483}, 3 (1997).  
\bibitem{tao} L. H. Tao {\it et al.}, Z. Phys. {\bf C70}, 387 (1996). 
\bibitem{tvaskis} V. Tvaskis, Ph.D. Thesis, Vrije Universiteit (2004), The Netherlands, unpublished.
\bibitem{bardin} A. A. Akhundov, D. Yu. Bardin and N. M. Shumeiko, Sov. J. Nucl. Phys. {\bf 26}, 660 (1977); 
D. Yu. Bardin and N. M. Shumeiko, Sov. J. Nucl. Phys. {\bf 29}, 499 (1979); and
A. A. Akhundov et al., Sov. J. Nucl. Phys. {\bf 44}, 988 (1986).
\bibitem{motsai} L. W. Mo and Y. S. Tsai, Rev. Mod. Phys. {\bf 41}, 205 (1969). 
\bibitem{f2all} H. Abramowicz and A. Levy, hep-ph/9712415, v1, (1997).
\bibitem{badelek} B. Badelek, J. Kwiecinski, and A. Stasto, Z. Phys. {\bf C74}, 297 (1997).


\end{thebibliography}
\end{document}